\documentclass[a4paper]{jpconf}
\usepackage{lineno,graphicx}
\DeclareUnicodeCharacter{2212}{\ensuremath{-}}
\begin{document}
\title{Hadronic resonance production with ALICE at the LHC}

\author{Sergey Kiselev$^*$ for the ALICE Collaboration}

%\address{$^*$NRC "Kurchatov Institute" - ITEP, B. Cheremushkinskaya 25, 117218 Moscow, Russia}
%\address{$^*$NRC $<<$Kurchatov institute$>>$, Moscow, Russia}
\address{$^*$NRC ``Kurchatov institute'', Moscow, Russia}

\ead{Sergey.Kiselev@cern.ch}

%\linenumbers
\begin{abstract}
Recent results on short-lived hadronic resonances obtained with the ALICE detector at LHC energies are presented. 
These results include  system-size  and  collision-energy evolution of transverse momentum spectra, 
yields and the ratios of resonance yields to those of longer-lived particles, and nuclear modification factors. 
Results will be compared with model predictions.
% and measurements at lower energies.
\end{abstract}

The study of resonance production is important in proton-proton, proton-nucleus, and heavy-ion collisions.
Since lifetimes of short-lived resonances are comparable with the lifetime of the late hadronic phase 
produced in heavy-ion collisions, resonance yields are affected by the regeneration and rescattering 
of their decay daughters in the hadronic phase. These competing effects are investigated by measuring the yield ratios 
of resonances to that of the ground state longer-lived hadron as a function of charged-particle multiplicity. 
From these measurements, it is possible to obtain information on the time interval between the chemical and the kinetic freeze-out.
Measurements in pp and p--Pb collisions constitute a reference for nuclear collisions 
and provide information for tuning event generators inspired by the Quantum Chromodynamics.
Moreover, some heavy-ion effects (elliptic flow~\cite{ALICEflow}, strangeness enhancement~\cite{ALICEstrenh}, ...) were also unexpectedly observed 
in small collision systems.

Results on short-lived 
mesonic $\rho(770)^{0}$, $\mathrm{K}^{*}(892)^{0}$, $\mathrm{K}^{*}(892)^{\pm}$, $f_{0}(980)$, $\phi(1020)$ 
as well as baryonic $\Sigma(1385)^{\pm}$, $\Lambda(1520)$ and $\Xi(1530)^{0}$ resonances 
(hereafter $\rho^{0}$, $\mathrm{K}^{*0}$, $\mathrm{K}^{*\pm}$, $f_{0}$, $\phi$, $\Sigma^{*\pm}$, $\Lambda^{*}$, $\Xi^{*0}$) 
have been obtained using data reconstructed with the ALICE detector.
The  resonances  are  reconstructed  in  their  hadronic  decay  channels  and have very different lifetimes
as shown in Table~\ref{tab:Res}. 
\begin{table}[ht]
\caption{Reconstructed decay mode, lifetime values~\cite{PDG} and the corresponding references where ALICE results for the hadronic resonances are reported.}
% and ALICE papers for hadronic resonances.}
\begin{center}
%\begin{tabular}{|c|c|c|c|c|c|}
%\begin{tabular}{ c c c c c c }
%\begin{tabular}{ l l l l l l l }
\begin{tabular}{ l l l l l l l l l}
%\hline
\br
%                    &$\rho^{0}$&$\mathrm{K}^{*0}$  &  $\Sigma^{*\pm}$ &    $\Lambda^{*}$    &     $\Xi^{*0}$   &  $\phi$    \\
                    &$\rho^{0}$&$\mathrm{K}^{*0}$  &$\mathrm{K}^{*\pm}$  & $f_{0}$ &  $\Sigma^{*\pm}$ &    $\Lambda^{*}$    &     $\Xi^{*0}$   &  $\phi$    \\
%\hline
\mr
%decay channel (B.R.)&$\pi\pi (1.00) $&$\mathrm{K}\pi (0.67) $&$\Lambda\pi (0.87) $&$p\mathrm{K} (0.22)$&$\Xi\pi (0.67)$&$\mathrm{K}\mathrm{K} (0.49) $\\
decay channel&$\pi\pi $&$\mathrm{K}\pi $&$\mathrm{K^{0}_{S}}\pi $&$\pi\pi $&$\Lambda\pi $&$p\mathrm{K} $&$\Xi\pi $&$\mathrm{K}\mathrm{K} $\\
%\hline
lifetime (fm/\it{c})& 1.3 & 4.2 & 4.2 & $\sim$ 5& 5-5.5 & 12.6 & 21.7& 46.2 \\ 
ALICE papers &\cite{ALICErho0}&\cite{ALICEpp7}-\cite{ALICEppXe} &\cite{ALICEppKstar-pm}-\cite{ALICEPb5Kstar-pm} &\cite{ALICEppf0}-\cite{ALICEpPbf0} &\cite{ALICEppSigmaStar}-\cite{ALICEPbPbSigmaStar} &\cite{ALICELambdaStar}-\cite{ALICELambdaStar2} &\cite{ALICEppSigmaStar}-\cite{ALICEpp13SstarXstar} &\cite{ALICEpp7}-\cite{ALICEpPb502},\cite{ALICEphiXe} \\
%systems             &   pp   &   pp   &   pp   &   pp   &   pp   &   pp   &   pp   &   pp \\
%                    &        &  p--Pb &        &        &  p--Pb &  p--Pb &  p--Pb &  p--Pb \\
%                    & Pb--Pb & Pb--Pb &        &        & Pb--Pb & Pb--Pb & Pb--Pb & Pb--Pb \\
%\hline
\br
\end{tabular}
\end{center}
 \label{tab:Res}
\end{table}
The V0A and V0C detectors (32 scintillating counters each) were used for the determination of 
the multiplicity classes by measuring the sum of the signals from V0A and V0C forming the V0M signal.
 
This contribution reports recent results obtained
for $\mathrm{K}^{*0}$ in Xe--Xe at $\sqrt{s_{\mathrm {NN}}}$ =5.44 TeV and in pp at 5.02 TeV \cite{ALICEppXe},
%for $\mathrm{K}^{*\pm}$ in pp at 13 TeV \cite{ALICEpp13Kstar-pm} and Pb--Pb at 5.02 TeV \cite{ALICEPb5Kstar-pm}, 
for $\mathrm{K}^{*\pm}$ in pp at 13 TeV and Pb--Pb at 5.02 TeV \cite{ALICEPb5Kstar-pm}, 
for $f_{0}$ in p--Pb at 5.02 TeV \cite{ALICEpPbf0},
for $\Sigma^{*\pm}$ and $\Xi^{*0}$ in pp at 13 TeV \cite{ALICEpp13SstarXstar}.

\begin{figure}[hbtp]
\begin{center}
\includegraphics[scale=0.35]{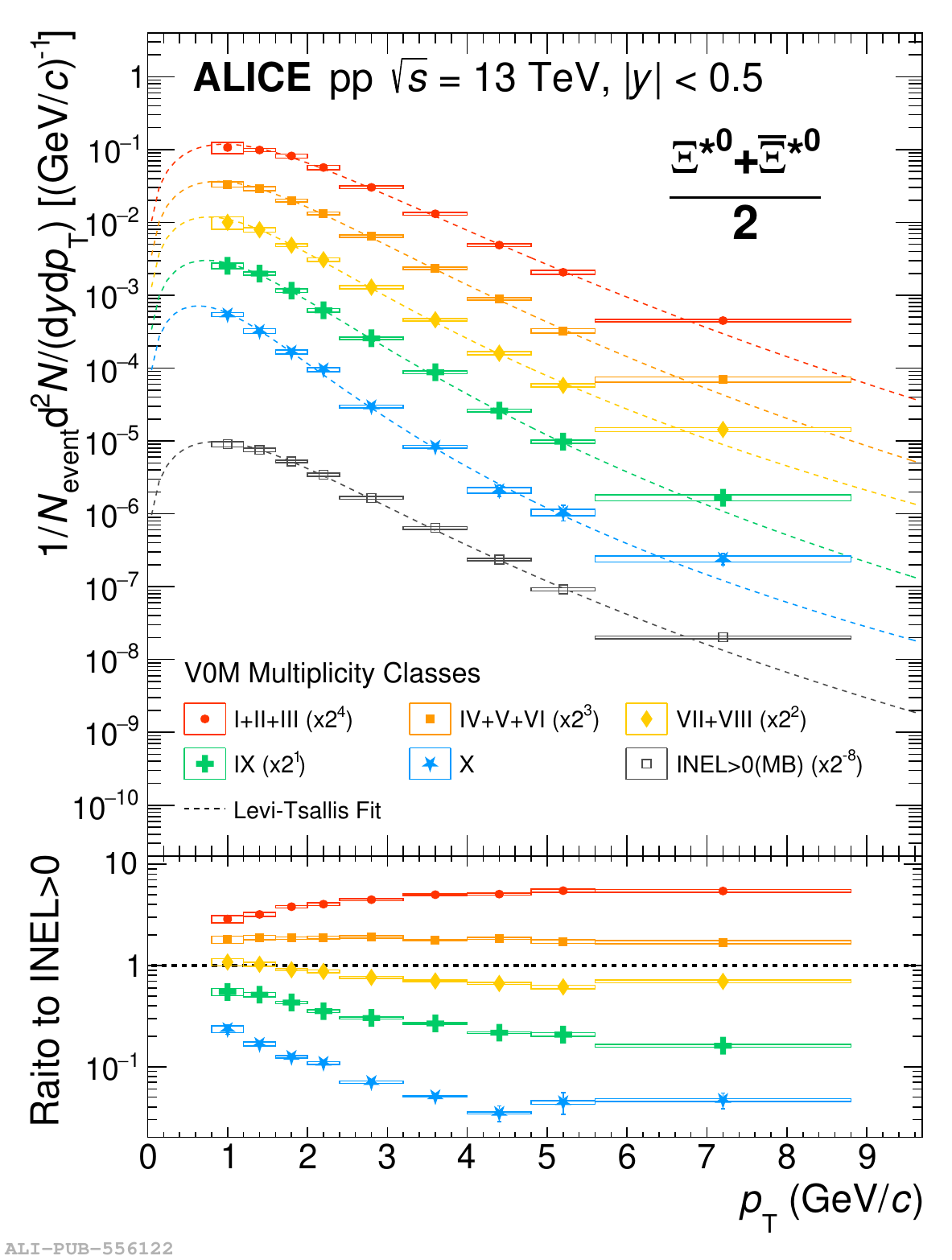}
\includegraphics[scale=0.44]{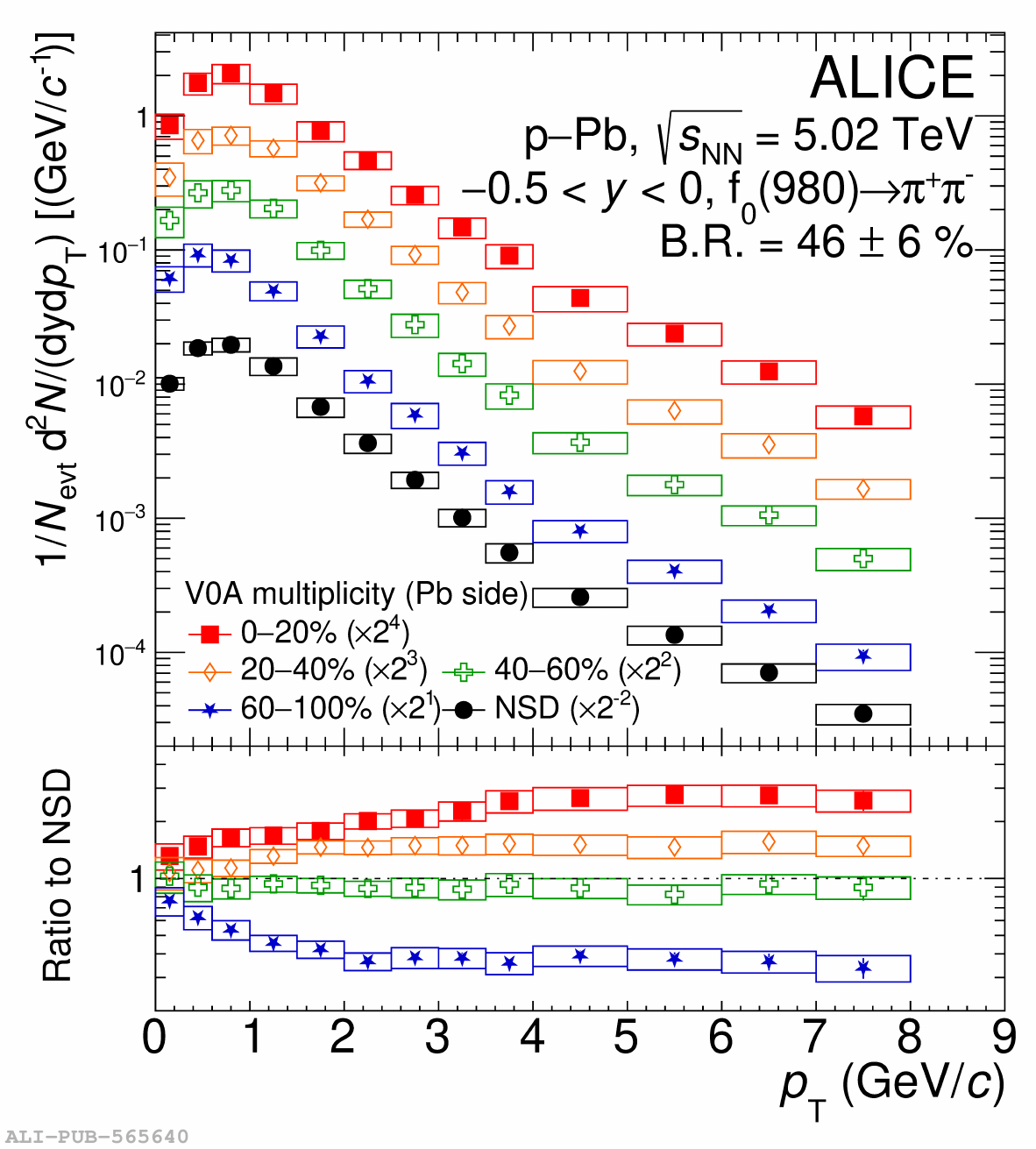}
\end{center}
\caption{(color online) 
Transverse momentum spectra of $\Xi^{*0}$ in pp collisions at $\sqrt{s}$ = 13 TeV (left) and $f_{0}$ in p–Pb collisions 
at $\sqrt{s_{\mathrm {NN}}}$ = 5.02 TeV (right) for different multiplicity classes.
}
  \label{fig:spectra}
\end{figure}

Figure~\ref{fig:spectra} shows the transverse momentum spectra of $\Xi^{*0}$ in pp collisions at $\sqrt{s}$ = 13 TeV and $f_{0}$ in p–Pb collisions
at $\sqrt{s_{\mathrm {NN}}}$ = 5.02 TeV for different multiplicity classes.
For $p_\mathrm{T} < 4$ GeV/$c$, a hardening of the $p_\mathrm{T}$ spectrum from low- to high-multiplicity events is clearly seen, 
while the spectral shapes in the different multiplicity classes are found to become consistent among each other for $p_\mathrm{T} > 4$ GeV/$c$. 
Such trends are similar to those observed for other hadron species \cite{ALICEpp13piKp}.
% and are understood as due to the radial flow.

\begin{figure}[hbtp]
\begin{center}
\includegraphics[scale=0.8]{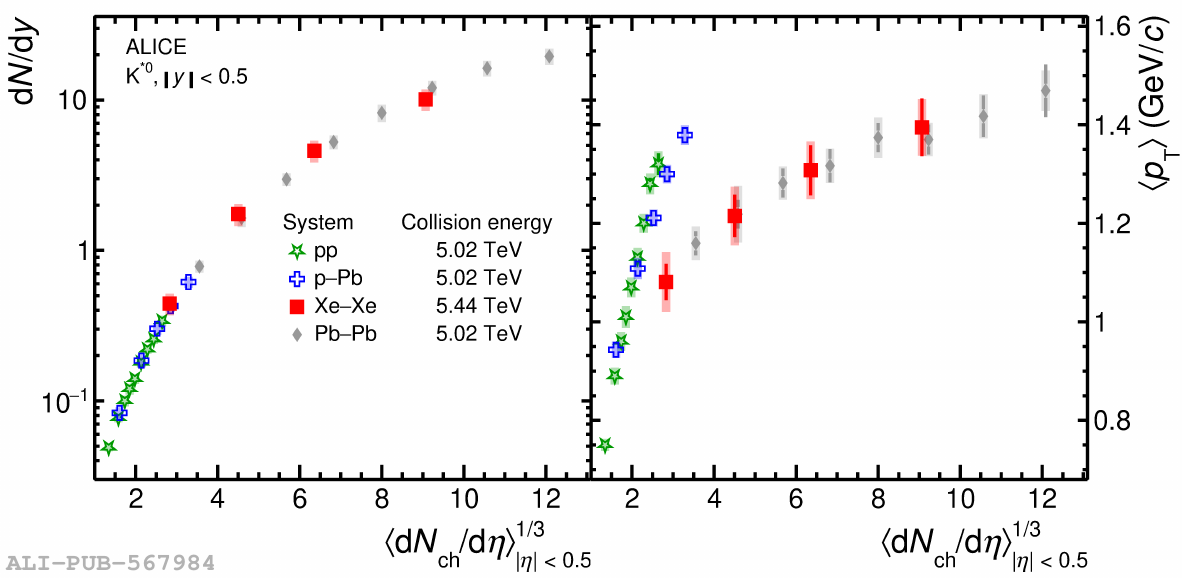}
\end{center}
\caption{(color online) 
The d$N$/d$y$ (left) and $\langle p_\mathrm{T}\rangle$ (right) of $\mathrm{K}^{*0}$ as a function of the charged-particle multiplicity density
in pp collisions at $\sqrt{s}$ = 5.02 TeV and Xe--Xe collisions at $\sqrt{s_{\mathrm {NN}}}$ = 5.44 TeV.
Measurements are compared with the results obtained in p--Pb~\cite{ALICEpPb} and Pb--Pb~\cite{ALICEppPbPb502} collisions at $\sqrt{s_{\mathrm {NN}}}$ = 5.02 TeV.
}
  \label{fig:mpt}
\end{figure}

Figure~\ref{fig:mpt} presents the d$N$/d$y$ and $\langle p_\mathrm{T}\rangle$ of $\mathrm{K}^{*0}$ as a function of the charged-particle multiplicity density.
For the d$N$/d$y$ new data for pp at $\sqrt{s}$ = 5.02 TeV and Xe--Xe at $\sqrt{s_{\mathrm {NN}}}$ = 5.44 TeV
follow a general trend: yields are independent of collision system and appear to be driven by the event multiplicity.
The $\langle p_\mathrm{T}\rangle$ rises faster with multiplicity in pp and p--Pb collisions than in Xe--Xe and Pb--Pb collisions.
% suggesting more rapid expansion.
An analogous behavior has been observed in~\cite{ALICE_mpt} for charged particles and can be understood 
as the effect of color reconnection between strings produced in multi-parton interactions.

\begin{figure}[hbtp]
\begin{center}
\includegraphics[scale=0.8]{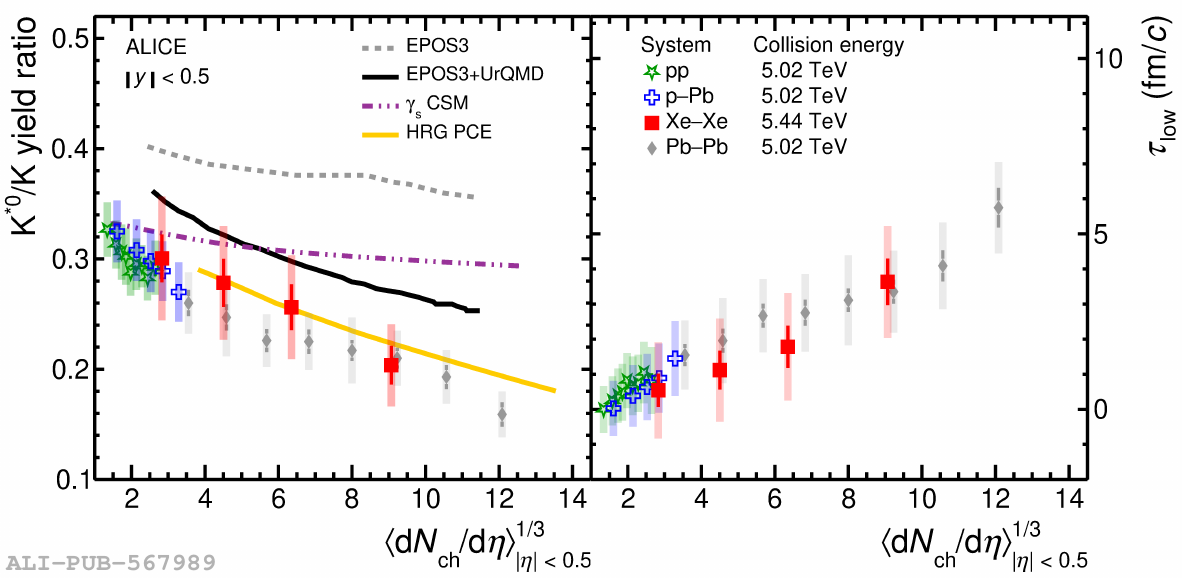}
\end{center}
\caption{(color online)
The particle yield ratio $\mathrm{K}^{*0}/\mathrm{K}$ (left) and the lower limit of hadronic phase lifetime (right) as a function of 
the charged-particle multiplicity density in pp collisions at $\sqrt{s}$ = 5.02 TeV and Xe--Xe collisions at $\sqrt{s_{\mathrm {NN}}}$ = 5.44 TeV.
Measurements are compared with the results obtained in p--Pb~\cite{ALICEpPb} and Pb--Pb~\cite{ALICEppPbPb502} collisions at $\sqrt{s_{\mathrm {NN}}}$ = 5.02 TeV.
For the ratio model predictions from EPOS3~\cite{KnospeEPOS} with and without the UrQMD afterburner, $\gamma_{S}$-CSM \cite{GsCSM} 
and HRG-PCE \cite{HRG-PCE} are also shown.
}
  \label{fig:Mratios}
\end{figure}

Figure~\ref{fig:Mratios} (left) shows the particle yield ratio $\mathrm{K}^{*0}/\mathrm{K}$ as a function of the charged-particle multiplicity density.
The yield ratio in different collision systems shows a smooth evolution with multiplicity, and is independent of the collision system.
Observed decrease in the ratio can be understood as the rescattering of $\mathrm{K}^{*0}$ meson’s decay daughters inside the hadronic phase~\cite{ALICEPbPb502}.
Data are compared with  EPOS3~\cite{KnospeEPOS}, $\gamma_{S}$-CSM~\cite{GsCSM} and HRG-PCE~\cite{HRG-PCE} predictions: HRG-PCE gives the best agreement with the data,
EPOS3 with the UrQMD afterburner qualitatively reproduces the multiplicity dependence, $\gamma_{S}$-CSM (rescattering effect is not implemented) does not explain the multiplicity dependence.
Using the measured ratios and the assumption that there is no regeneration of $\mathrm{K}^{*0}$ in the hadronic medium,
one can obtain an estimate of the lower bound of the hadronic phase lifetime $\tau_{\rm low}$, i.e., the time between chemical and
kinetic freeze-out,~Fig.~\ref{fig:Mratios} (right). The $\tau_{\rm low}$ smoothly increases with multiplicity and reaches $\sim$ 5 fm/$c$.

\begin{figure}[hbtp]
\begin{center}
\includegraphics[scale=0.44]{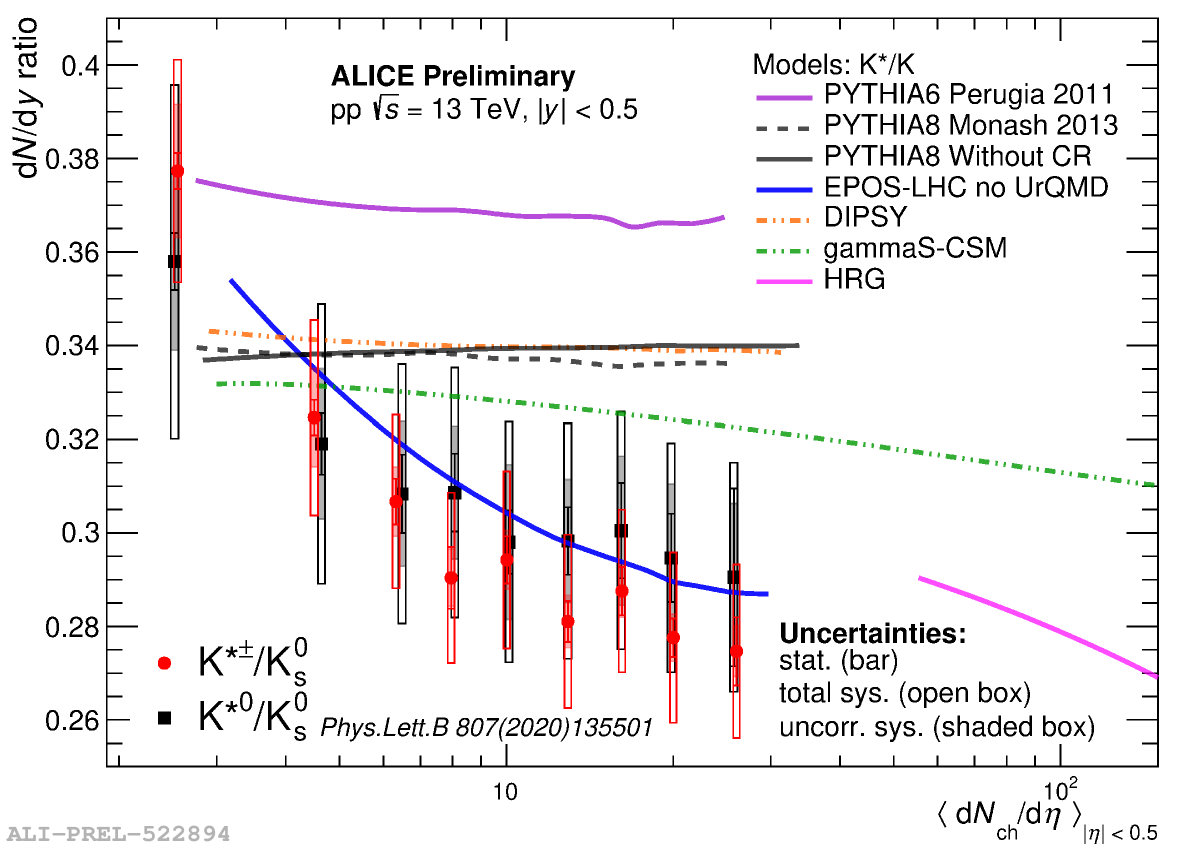}
\includegraphics[scale=0.33]{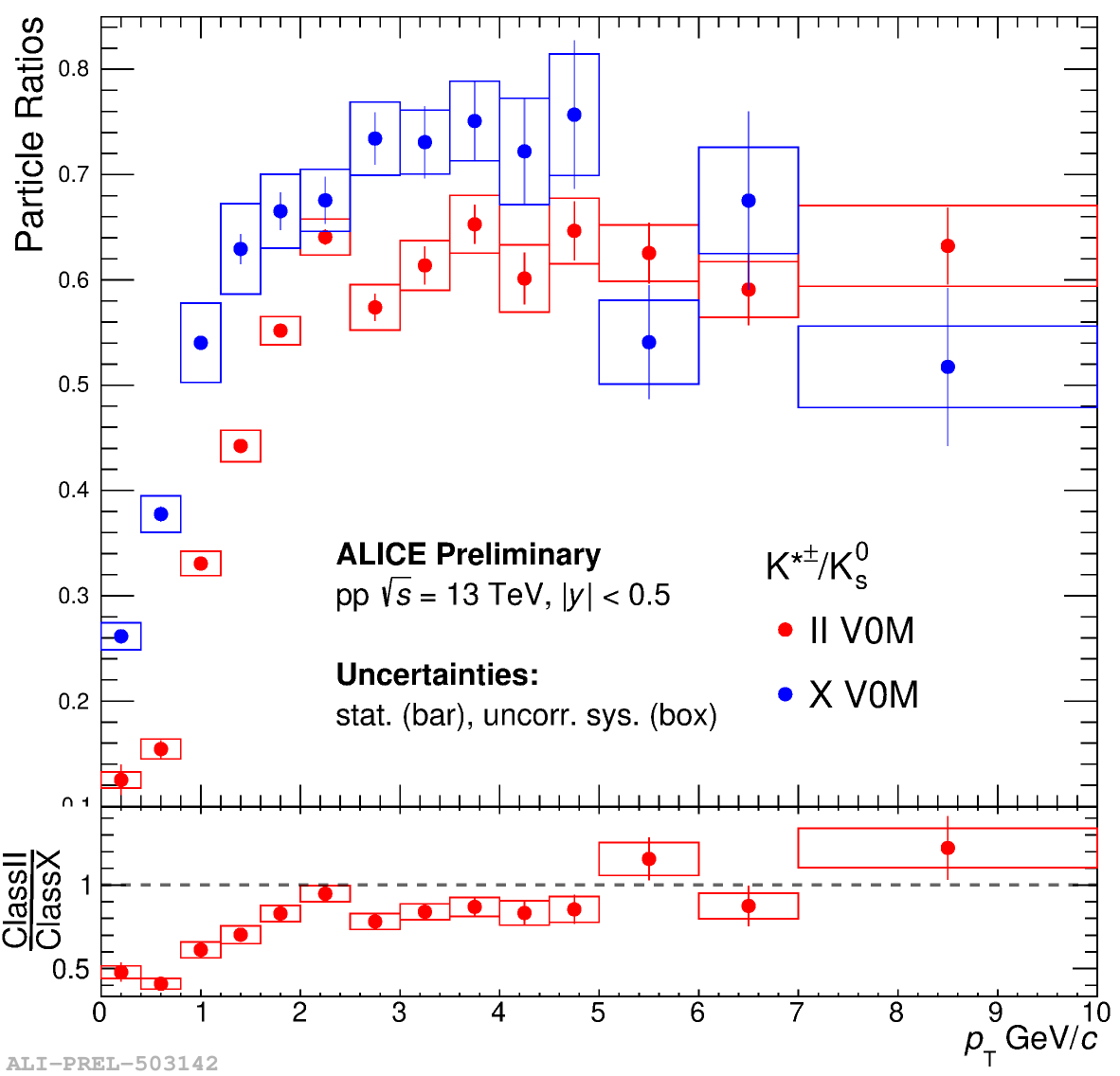}
\end{center}
\caption{(color online)
Left: Ratios $\mathrm{K}^{*\pm}/\mathrm{K}$ and $\mathrm{K}^{*0}/\mathrm{K}$~\cite{ALICEpp13} as a function of the charged-particle multiplicity density
in pp collisions at $\sqrt{s}$ = 13 TeV.
Model predictions from PYTHIA6 \cite{Perugia}, PYTHIA8 \cite{Monash}, EPOS-LHC \cite{EPOSLHC}, DIPSY \cite{DIPSY}, $\gamma_{S}$-CSM (gammaS-CSM) \cite{GsCSM} 
and HRG \cite{HRG-PCE} are also shown. 
Right: Ratios $\mathrm{K}^{*\pm}/\mathrm{K}$ as a $p_\mathrm{T}$ function for low (X) and high (II) multiplicity classes. 
}
  \label{fig:Kratio13}
\end{figure}

Figure~\ref{fig:Kratio13} (left) shows the the multiplicity dependence of the $\mathrm{K}^{*\pm}/\mathrm{K}$ ratio in pp collisions at $\sqrt{s}$ = 13 TeV, 
compared to the $\mathrm{K}^{*0}/\mathrm{K}$ one.
The decreasing trend already outlined by the $\mathrm{K}^{*0}$ analysis is confirmed by the $\mathrm{K}^{*\pm}$ results.
The $\mathrm{K}^{*\pm}/\mathrm{K}$ ratio in the highest multiplicity class is below the low multiplicity value at $\sim$ 7$\sigma$ level 
taking into account the multiplicity uncorrelated uncertainties ($\sim$ 2$\sigma$ level for the $\mathrm{K}^{*0}/\mathrm{K}$ ratio).
This result represents the first evidence of a clear $\mathrm{K}^{*}/\mathrm{K}$ suppression measured in small collision systems.
EPOS-LHC \cite{EPOSLHC} provides good agreement with the measured data, well reproducing the decreasing trend, while PYTHIA6 \cite{Perugia}, PYTHIA8 \cite{Monash},
 DIPSY \cite{DIPSY} and $\gamma_{S}$-CSM tend to overestimate the ratios at high multiplicities and exhibit a fairly flat trend.
The ratio of the high multiplicity $\mathrm{K}^{*\pm}/\mathrm{K}$ differential $p_\mathrm{T}$ distribution to the low multiplicity one helps 
to quantify the observed decrease in the particle ratios, Fig.~\ref{fig:Kratio13} (right). For $p_\mathrm{T} \leq$ 2 GeV/$c$ the  double $\mathrm{K}^{*\pm}/\mathrm{K}$ ratio 
deviates from unity by more than 3$\sigma$ suggesting a low $p_\mathrm{T}$ dominant process.

\begin{figure}[hbtp]
\begin{center}
\includegraphics[scale=0.8]{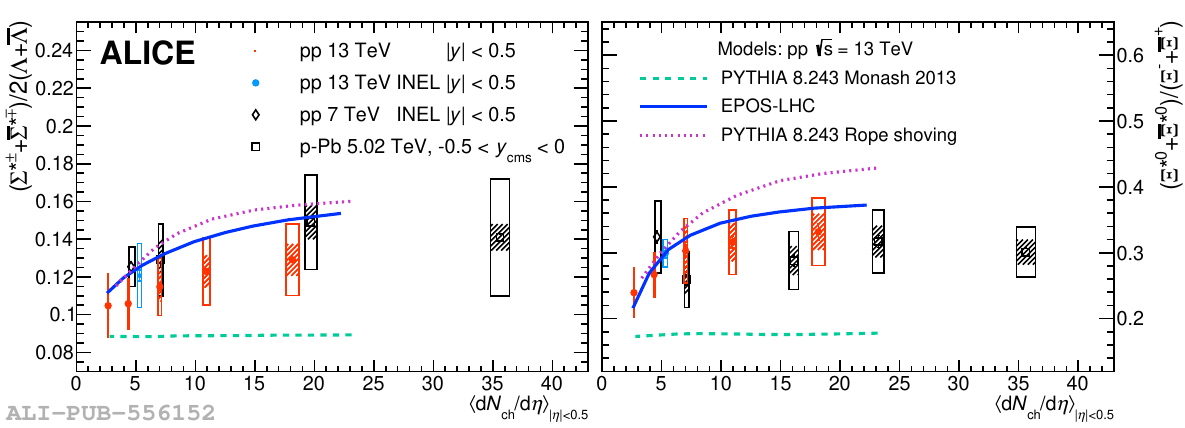}
\end{center}
\caption{(color online) 
Ratios $\Sigma^{*\pm}/\Lambda$ (left) and $\Xi^{*0}/\Xi$ (right) as a function of the charged-particle multiplicity density in pp collisions at $\sqrt{s}$ = 13 TeV.
Measurements are compared with the results obtained in pp at $\sqrt{s}$ = 7 TeV~\cite{ALICEppSigmaStar} 
and p--Pb at $\sqrt{s_{\mathrm {NN}}}$ = 5.02 TeV~\cite{ALICEpPbSigmaStar}.
Model predictions from EPOS-LHC \cite{EPOSLHC}, PYTHIA8 \cite{Monash} and PYTHIA8 with Rope shoving \cite{PYTHIA8shoving} are also shown. 
}
  \label{fig:SXratio13}
\end{figure}

Ratios $\Sigma^{*\pm}/\Lambda$ (left) and $\Xi^{*0}/\Xi$ (right) are illustrated in Fig.~\ref{fig:SXratio13}. 
In the new results for pp collisions at $\sqrt{s}$ = 13 TeV there is a hint of increase of the ratio with increasing multiplicity.
Despite similar lifetimes, $\mathrm{K}^{*}$ and $\Sigma^{*\pm}$ exhibit different trends.
One of decay daughters ($\Sigma^{*\pm} \rightarrow \Lambda\pi$, $\Xi^{*0} \rightarrow \Xi\pi$) are long-lived particles $\Lambda$ and $\Xi$, which decay out of the hadron phase.
Only $\pi$ can be rescattered and the effect of regeneration $\Lambda\pi \rightarrow \Sigma^{*\pm}$ can be more pronounced.
For the longer-lived $\Xi^{*0}$, $c\tau$=21.7, the hadronic stage effects are less important.
The EPOS-LHC and PYTHIA8 with Rope shoving predict a slight increase of the ratios.

\begin{figure}[hbtp]
\begin{center}
\includegraphics[scale=0.8]{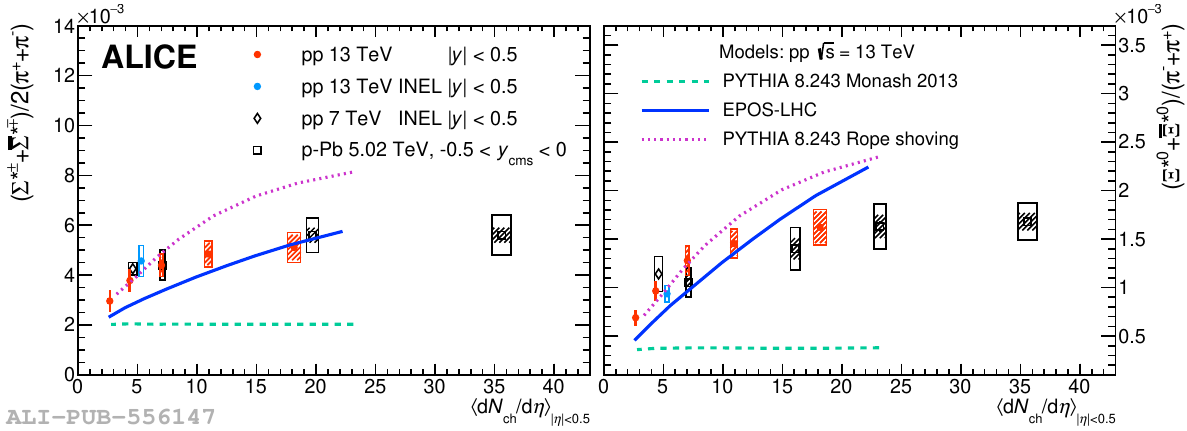}
\end{center}
\caption{(color online) 
Ratios $\Sigma^{*\pm}/\pi$ (left) and $\Xi^{*0}/\pi$ (right) as a function of the charged-particle multiplicity density in pp collisions at $\sqrt{s}$ = 13 TeV.
Measurements are compared with the results obtained in pp at $\sqrt{s}$ = 7 TeV~\cite{ALICEppSigmaStar} 
and p--Pb at $\sqrt{s_{\mathrm {NN}}}$ = 5.02 TeV~\cite{ALICEpPbSigmaStar}.
Model predictions from EPOS-LHC \cite{EPOSLHC}, PYTHIA8 \cite{Monash} and PYTHIA8 with 
Rope shoving \cite{PYTHIA8shoving} are also shown. 
}
  \label{fig:SXratiopi13}
\end{figure}

Figure~\ref{fig:SXratiopi13} shows ratios $\Sigma^{*\pm}/\pi$ (left) and $\Xi^{*0}/\pi$ (right). 
The results show a smooth increasing trend as a function of multiplicity without energy and collision system
dependence. The increase depends on the strangeness content of the resonance; with $\Sigma^{*\pm}$ having 
a strangeness content (S) of 1 and $\Xi^{*0}$ having S=2. These results are consistent with previous 
measurements of ground-state hyperons to pion ratios with ALICE~\cite{ALICEpp13multDepS}. 
EPOS-LHC and PYTHIA8 with Rope shoving predict an increasing trend with multiplicity for both resonances.

\begin{figure}[hbtp]
\begin{center}
\includegraphics[scale=0.40]{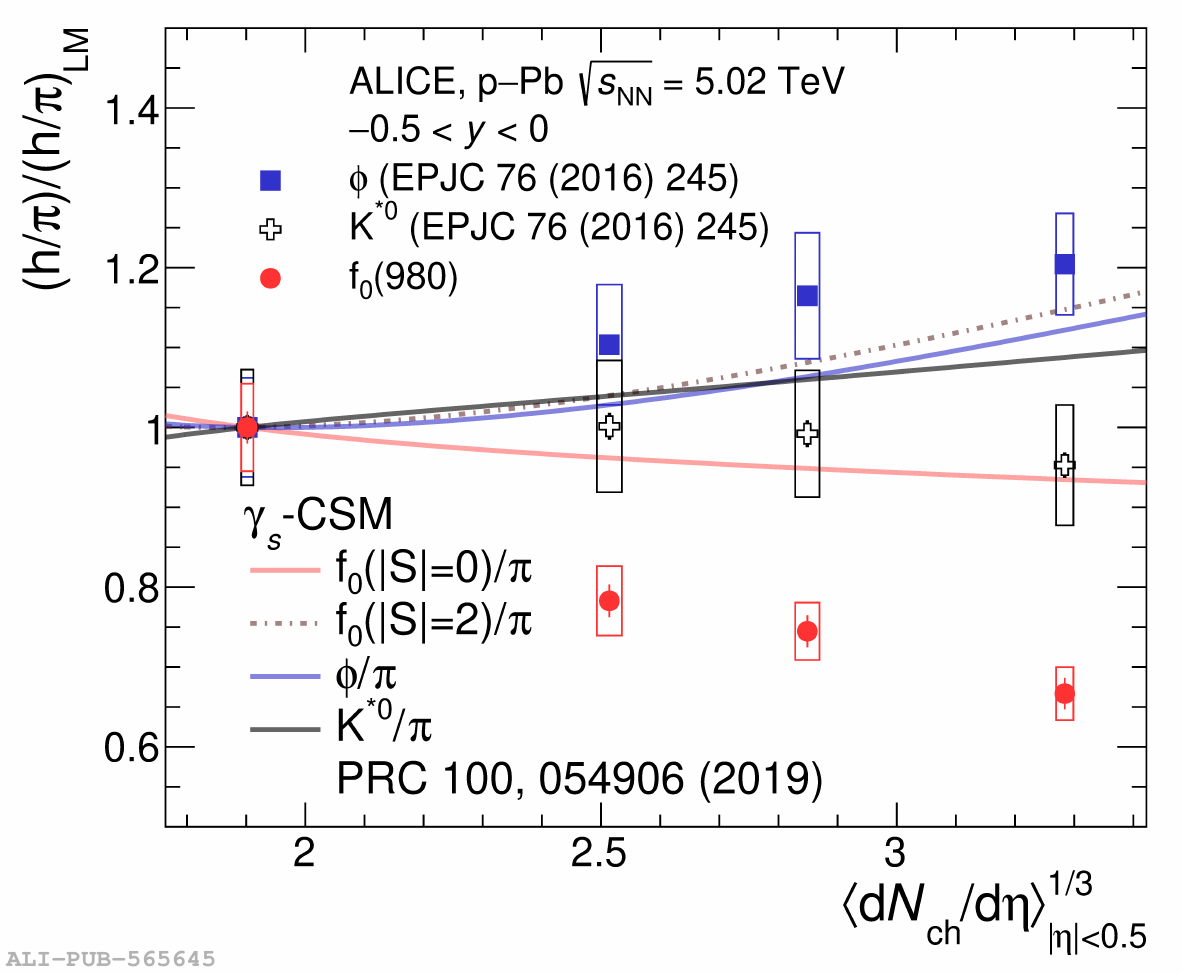}
\includegraphics[scale=0.37]{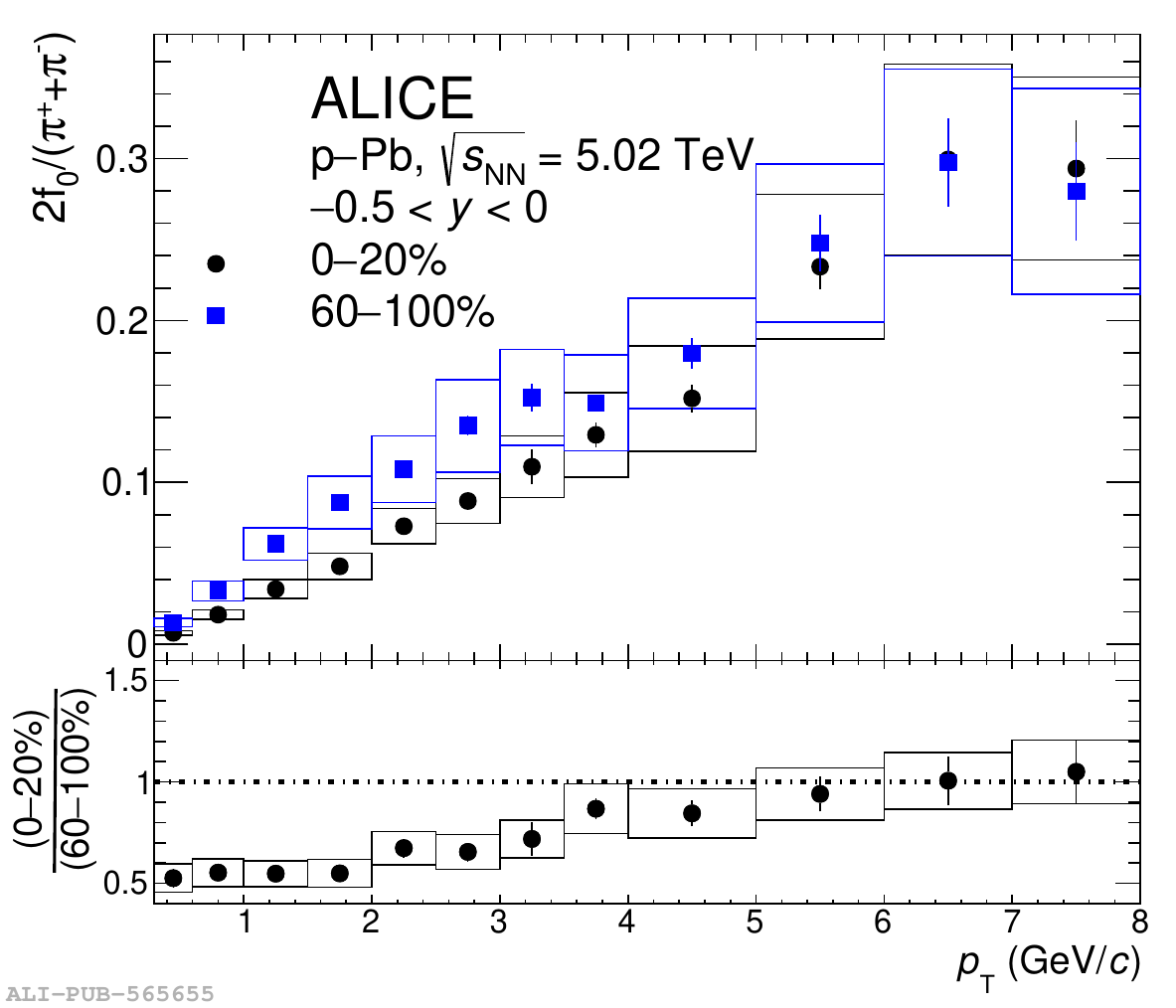}
\end{center}
\caption{(color online)
Left: Double ratio of particle yields to pion yields as a function of multiplicity in p--Pb collisions at $\sqrt{s_\mathrm{NN}}$  = 5.02 TeV.
Model predictions from $\gamma_{S}$-CSM \cite{GsCSM} are also shown.
Right: $p_\mathrm{T}$-differential particle yield ratio of $f_{0}$ to pion. 
}
  \label{fig:f0}
\end{figure}

%Quark structure of $f_{0}(980)$ is still debated and among possible configurations there are 
%$q\overline{q}$, $(q)^{2}(\overline{q})^{2}$, hadronic molecules,...~\cite{ALICEppf0}.
To investigate the structure of $f_{0}$, the new ALICE measurements of the $f_{0}$ yields
are compared to those of other hadrons and resonances with similar mass.
Figure~\ref{fig:f0} (left) shows the double ratio of particle yields to pion yields as a function of multiplicity 
in p--Pb collisions at $\sqrt{s_\mathrm{NN}}$  = 5.02 TeV.
Strangeness enhancement with multiplicity could explain the increase with multiplicity of the $\phi/\pi$ ratios.
The $\mathrm{K}^{*0}/\pi$ ratios demonstrate the competition between strangeness
enhancement and rescattering effect. 
The $f_{0}/\pi$ suppression shows that rescattering is the effect that dominantly affects 
the yield at low $p_{\rm T}$, Fig.~\ref{fig:f0} (right). 
The $p_{\rm T}$ -differential $f_{0}/\pi$ ratio does not exhibit the characteristic enhancement of baryon-to-meson ratios, 
suggesting a structure with two constituent quarks for the $f_{0}$ resonance.
The $\gamma_{S}$-CSM model qualitatively reproduces the $\phi/\pi$ ratio
and overestimates the $\mathrm{K}^{*0}/\pi$ ratio. For the $f_{0}/\pi$ ratio predictions with zero hidden strangeness, $|S|$=0,
are closer to the data values.
%We have a hint of zero net strangeness of $f_{0}$.
%Notably, the $\gamma_{S}-$CSM model prediction for the $f_{0}/\pi$ ratio assuming zero net strangeness of $f_{0}$ 
%is consistent with the data within 1.9$\sigma$ ~\cite{ALICEppf0}.

\begin{figure}[hbtp]
\begin{center}
\includegraphics[scale=0.8]{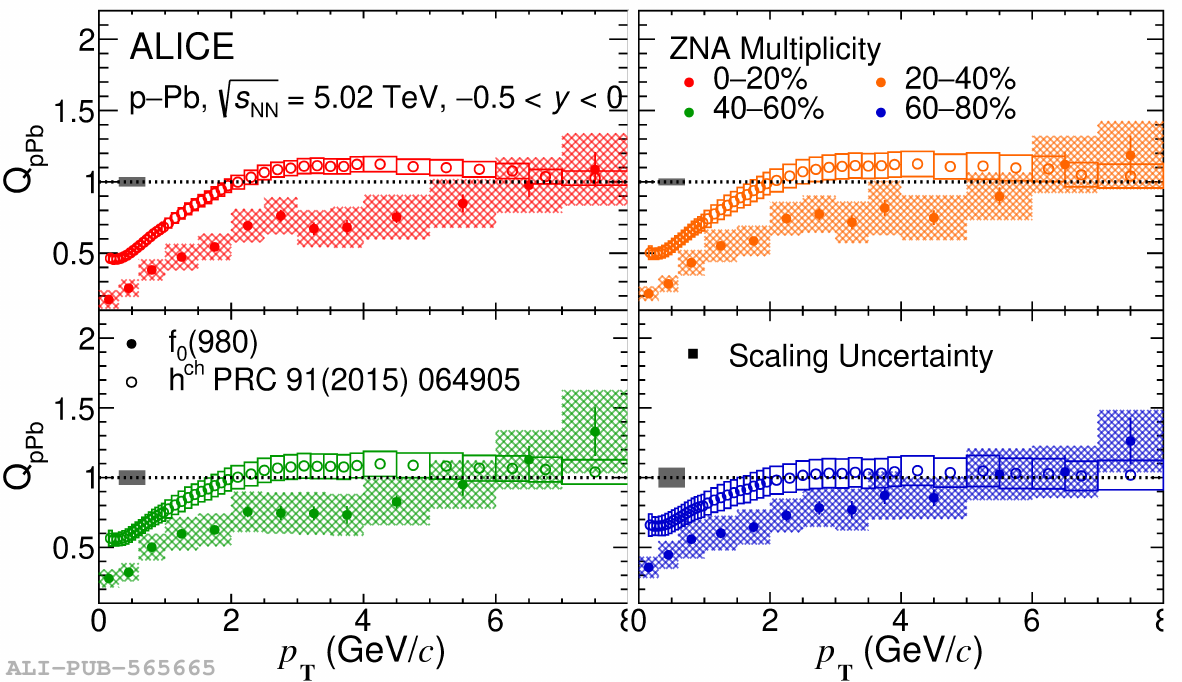}
\end{center}
\caption{(color online) 
The nuclear modification factor $Q_{\rm pPb}$ of $f_{0}$ as a  function of $p_\mathrm{T}$  in p--Pb collisions at $\sqrt{s_\mathrm{NN}}$  = 5.02 TeV
for different multiplicity classes. The $Q_{\rm pPb}$ of charged hadrons \cite{ALICEpPb5centr} are reported for comparison.
}
  \label{fig:QpPb}
\end{figure}

Figure~\ref{fig:QpPb} shows the nuclear modification factor $Q_{\rm pPb}$ of $f_{0}$ as a  function of $p_\mathrm{T}$  
in p--Pb collisions at $\sqrt{s_\mathrm{NN}}$  = 5.02 TeV for different multiplicity classes.
The $Q_{\rm pPb}$ of $f_{0}$ does not exhibit a Cronin-like enhancement.
Since baryons demonstrate a Cronin-like enhancement and mesons show little or no nuclear modification \cite{ALICEpPb816}, \cite{ALICEpPb5piKp}
might suggest that the $f_{0}$ is composed of two quarks.
At low $p_\mathrm{T} < $ 4 GeV/$c$ the suppression of $f_{0}$ becomes more pronounced with increasing multiplicity. This can be explained by the rescattering and radial flow effects.
 
In summary, recent results on short-lived hadronic resonances obtained by the ALICE experiment in pp, p--Pb, Xe--Xe and Pb--Pb collisions 
at the LHC energies have been presented.
% Fig.2a
Yields of $\mathrm{K}^{*0}$ are independent of the collision system and appear to be driven by the event multiplicity.
% Fig.2b
In pp and p--Pb collisions the $\langle p_\mathrm{T}\rangle$ values of $\mathrm{K}^{*0}$
rise faster with multiplicity than in Xe--Xe and Pb--Pb collisions.
One possible explanation could be the effect of color reconnection between strings produced in multi-parton interactions.
% Fig.3
The $\mathrm{K}^{*0}/\mathrm{K}$ ratio in different collision systems shows a smooth evolution with multiplicity, 
and is independent of the collision system. Observed decrease in the ratio can be understood as the rescattering 
of $\mathrm{K}^{*0}$ meson’s decay daughters inside the hadronic phase. 
Among the models, HRG-PCE gives the best agreement with the data. 
The lower bound of the hadronic phase lifetime estimated with these ratios smoothly increases with multiplicity and reaches $\sim$ 5 fm/c
at highest multiplicity.
% Fig.4
For pp collisions at $\sqrt{s}$ = 13 TeV the $\mathrm{K}^{*\pm}/\mathrm{K}$ ratio in the highest multiplicity class is below 
the low multiplicity value at $\sim$ 7$\sigma$ level representing the first evidence of a $\mathrm{K}^{*}/\mathrm{K}$ suppression measured in pp collisions.
 The EPOS-LHC model provides good agreement with the measured data.
% Fig.5, 6
For pp collisions at $\sqrt{s}$ = 13 TeV there is also a hint of increase of the $\Sigma^{*\pm}/\Lambda$ and $\Xi^{*0}/\Xi$ ratios with increasing multiplicity.
For $\Sigma^{*\pm}$ this may be an indication of the predominance of regeneration over rescattering.
The EPOS-LHC and PYTHIA8 with Rope shoving predict a slight increase of the ratios.
The $\Sigma^{*\pm}/\pi$ and $\Xi^{*0}/\pi$ ratios show a smooth increasing trend as a function of multiplicity
and are consistent with previous measurements of ground-state hyperons to pion ratios.
EPOS-LHC and PYTHIA 8 with Rope shoving predict an increasing trend with multiplicity for both resonances.
% Fig.7
In p--Pb collisions at $\sqrt{s_\mathrm{NN}}$  = 5.02 TeV the $f_{0}/\pi$ suppression shows that rescattering is the effect that dominantly affects 
the yield at low $p_{\rm T}$. 
The $p_{\rm T}$ -differential $f_{0}/\pi$ ratio does not exhibit the characteristic enhancement of baryon-to-meson ratios, 
suggesting a structure with two constituent quarks for the $f_{0}$ resonance.
% Fig.8
The nuclear modification factor $Q_{\rm pPb}$ of $f_{0}$ does not exhibit a Cronin-like enhancement
suggesting that the $f_{0}$ is composed of two quarks.

The work was carried out within the state assignment of NRC ``Kurchatov institute''.

\end{document}